\input harvmac.tex
\Title{\vbox{\baselineskip12pt\hbox{CTP-TAMU-44/96}\hbox{hep-th/9609048}}}
{\vbox{\centerline{A Note on Brans-Dicke Cosmology with Axion}}}
\centerline{\bf Sudipta Mukherji}
\smallskip\centerline{\it Center for Theoretical Physics, Department of 
Physics}
\smallskip\centerline{\it Texas A \& M University, College Station, Texas
77843-4242, USA}
\smallskip\centerline{e-mail: mukherji@bose.tamu.edu}

\vskip .3in
We study the Brans-Dicke model in the presence of an axion. The dynamical 
equations are solved when the fields are space independent and the 
metric is spatially flat. It is found that at late time the scale factor 
undergoes decelerated expansion but the dilaton grows large. At early 
time, scale factor and the dilaton approach constants.
\Date{9/96}
\def \bal{\bar \alpha}
\def \part{\partial}
\def \om {\omega}
\def \bp {{\bar \phi}}

\lref\venezianone{G. Veneziano, Phys. Lett. {\bf B265}, (1991) 287.}

\lref\gasv{M. Gasperini and G. Veneziano, Astropart. Phys. {\bf 1}, (1993) 
317.}

\lref\clw{E. Copeland, A. Lahiri and D. Wands, Phys. Rev. {\bf D50}, 
(1994) 4868.}

\lref\kalyan{S. K. Rama, {\it Singularity free (Homogeneous and 
Isotropic) Universe in Graviton-Dilaton Models}, hep-th/9608024.}

\lref\barrow{J. Barrow and P. Parsons, {\it The Behaviour of Cosmological 
Models with Varying G}, gr-qc/9607072.}

\lref\lidsey{J. Lidsey, {\it Scale Factor Duality and Hidden 
Supersymmetry in Scalar-Tensor Theories}, Phys. Rev {\bf D52}(1995) 5407.}

\lref\mw{J. Mimoso and D. Wands, {\it Massless Fields in Scalar-Tensor
Cosmologies}, Phys. Rev. {\bf D51} (1995) 477.}

\lref\veneziano{M. Gasperini, J. Maharana and G. Veneziano, {\it
Graceful Exit in Quantum String Cosmology}, hep-th/9602087.}

\lref\bv{R. Brustein and G. Veneziano, Phys. Lett {\bf B} 329, (1994), 
429.} 

\lref\tseytlin{A. Tseytlin, Int. J. Mod. Phys. {\bf D1} (1992) 223.}

\lref\gper{D. Goldwirth and M. Perry, Phys. Rev. {\bf D49}, (1994) 
5019.}  

\lref\emw{R. Easther, K. Maeda and D. Wands, {\it Tree-level String 
Cosmology}, hep-th/9509074.}

\lref\callan{C. Callan, D. Friedan, E. Martinec and M. Perry, Nucl. Phys. 
{\bf B} 262 (1985) 593.}

\lref\maeda{R. Easther and K. Maeda, {\it One-Loop Superstring Cosmology 
and the Non-Singular Universe}, hep-th/9605173.}

\lref\kmo{N. Kaloper, R. Madden and K. Olive, {\it Axions and the 
Graceful Exit Problem in String Cosmology}, hep-th/9510117.}

\lref\gasperini{M. Gasperini and G. Veneziano, {\it Birth of the Universe 
as Quantum Scattering in String Cosmology}, hep-th/9602096.}

\lref\antoniadis{I. Antoniadis, J. Rizos and K. Tamvakis, Nucl. Phys. 
{\bf B415} (1994) 497.}

\noindent {\bf 1.}
Exploring the implications of duality symmetry in string theory, 
the possibility of 
having inflation without relying on the potential energy density
is pointed out in \refs{\venezianone}, \refs{\gasv}, \refs{\bv}.
However, exiting from inflationary period turned out to be  a problem 
essentially because of the presence of singularities \refs{\bv}. 
Two possible 
directions have been proposed since then. First one is to investigate the 
higher order corrections in string action which might become strong to 
smooth out the singularity \refs{\antoniadis}, \refs{\maeda}, 
\refs{\kalyan}. The 
second  is the possibility of exiting from the inflationary period via 
quantum tunneling through classically forbidden region \refs{\veneziano}
, \refs{\gasperini}.

Besides  graviton and dilaton, string theory predicts a three 
rank tensor 
field($H$) in the massless spectrum of the theory. In \refs{\tseytlin},
\refs{\gper}, \refs{\clw}, \refs{\kmo}, \refs{\emw} cosmological 
evolutions have 
been studied is presence of such field. In particular, in \refs{\clw},
a general analytic solution has been given for flat 
Friedmann-Robertson-Walker(FRW) metric. 
It was shown that the presence of 
this $H$ field significantly modifies the evolution 
of the metric and the 
dilaton. It was found that the very presence of axion field may prevent 
the universe from a singular collapse.

On the other hand, recently in \refs{\lidsey}, exploiting a possible 
scale factor duality symmetry, Brans-Dicke(BD) theory 
in presence of the cosmological constant has been analyzed. 
Here, in the action the dilaton kinetic energy term carries a free 
parameter $\om$ instead of usual $\om = -1$ that appears in the low 
energy string theory(for more on these theories, even when $\om$ is a 
dilaton dependent function, see the 
paper \refs{\barrow} and references 
there in). If we also introduce a three 
rank tensor field in such model, 
the action takes the form
\eqn\action{S = \int d^4x{\sqrt {-g}}e^{-\phi}
(R - \omega \partial_\mu\phi\partial^\mu\phi - {1\over {12}} H^2 ),}
where $\phi$ is the dilaton field. It is not very clear if this
model with $\om \ne -1$ can arise in string theory but let us however 
note the following. String theory in four dimension will certainly 
contain many moduli fields. The detail of their number and structure 
follows from the detail of the 
compactification scheme. Nevertheless, if 
it so happens that one of the moduli fields has a flat direction along 
the dilaton, one might expect to have $\om$ different from $-1$. At
the same time,
however, even if this action is not derivable 
from string theory, it may 
be worthwhile to study it on its own right as an interesting model
for gravity theories. At this stage, we would like to mention that 
BD cosmology has also been studied \refs{\mw}
in the presence of a particular 
class of massless matter where the massless field does not couple to the 
dilaton in the BD frame (unlike the 
action \action\ here). These models may 
arise in string 
theory where the field strength (analogue of $H$ here) comes from the 
Ramond-Ramond sector rather than the 
Neveu-Schwarz sector as it is in our 
case. 

 In what follows, we  study \action\ ~in a flat FRW background.
\bigskip
\noindent{\bf 2.}
We take the metric of the form:
\eqn\metric{dS^2 = -dt^2 + e^{2\alpha(t)}dx^idx^j \delta_{ij}.} 
Provided all the fields are independent of the space coordinates,
the action \action\ can be brought to the following form:
\eqn\newac{S = \int dt e^{(3\alpha -\phi)}\{
-6(\partial_t \alpha)^2 + 6\partial_t\alpha \partial_t\phi +
\om(\partial_t\phi)^2-{q^2\over 2}e^{-6\alpha}\}.}
The constant $q$ is related to the three rank tensor fields 
\refs{\emw} ~as follows: 
the dual of $H_{\mu\nu\rho}$ is an axion 
($\Theta$)in four dimension which is defined as
\eqn\psu{H^{\mu\nu\rho} = e^\phi 
\epsilon^{\mu\nu\rho\lambda}\partial_\lambda \Theta.}
Since the action does not have any potential 
energy contribution for the 
axion field, the momentum follows from the above 
lagrangian is constant. So
we have defined $p_\Theta = {\partial{\cal 
L}\over{\partial\partial_t\Theta}} =q$.

The constraint equation and the equations of motion are:
\eqn\one{
\om\part_t^2\phi + 3(1+\om)({\part_t a\over a})\part_t\phi + 3
({\part_t^2 a\over a}) =0,}
\eqn\two{
6({\part_t a\over a})^2 - 6({\part_t a\over a})\part_t\phi 
-\om(\part_t\phi)^2 - {q^2\over 2}e^{-6\alpha} =0,}
\eqn\three{
\om \part_t^2\phi -6({\part_t a\over a})^2 + 3(\om +2)
({\part_t a\over a})\part_t\phi + 
{3q^2\over {2(2\om +3)}}e^{-6\alpha} =0.}
where the scale factor $a = e^\alpha$.

Fortunately \one\ - \three\ can be solved exactly for 
generic $\om$($> -{3\over 2}$).
The solutions are (the detail is given in the appendix)
\eqn\four{
\phi = -{1\over 2}{\rm ln}[{c(2\om +3)\over {4q^2}}{\rm sech}^2{
\sqrt c\tau\over 2}],}
\eqn\five{
a = [{4q^2\over 
{c(2\om +3)}}{\rm cosh}^2{\sqrt c\tau\over 2}]^{1\over 4}
e^{{1\over 4}{\sqrt{c(2\om +3)\over 3}\tau}},}
where the dilaton-time $\tau$ is defined as 
\eqn\six{t = \int e^{-\bp}d\tau}
with
\eqn\seven{
\bp = {1\over 4}[{\rm ln}\{{c(2\om +3)\over {4q^2}}{\rm sech}^2{
\sqrt c\tau\over 2}\} -\sqrt{3c(2\om +3)}\tau].}
Here $c$ is an arbitrary constant. All other constants that appear 
from integrating \one\ - \three\ are set to zero 
for convenience. Since it 
will not play an important 
role in our discussion, for simplicity 
we will also set $c = {4q^2\over(2\om 
+3)}$. Now using \six\ , one can express \four\ , \five\ in terms of
cosmic time $t$. Though, \six\ is hard to compute exactly,
it can be integrated numerically. The behaviour is shown in fig.1 for
$\om =1$ and $q^2 =5$.
On the other hand, it is easy to extract out early 
and late time behaviour of the scale factor and the dilaton.
 It turns out that for 
$\tau\rightarrow\infty$(or $t\rightarrow\infty$),
\eqn\nine{
e^\phi~\sim~t^{2\over {1 + \sqrt{3(2\om +3)}}}, }
\eqn\eight{
a~\sim~t^p,~~~p = {{1 + \sqrt{2\om +3\over 3}}\over {1 + 
\sqrt{3(2\om +3)}}},~~{\rm with}~{da\over dt} > 0,~~{d^2a\over {dt^2}} < 
0.}
So we see that the scale 
factor undergoes a decelerated expansion similar 
to the post-big bang regime as in \refs{\venezianone}, 
\refs{\gasv}. However, the theory becomes 
strongly coupled at late time as 
the original coupling constant $e^\phi$ 
gets large. It is easy to check 
that for $\om = -1$, we recover the behaviour 
of \refs{\gper}, \refs{\clw}. 
Furthermore, when $\om \rightarrow \infty$, we have 
$a \sim t^{1\over 3}$ and $e^\phi$ goes to a constant.
When $\tau\rightarrow 0$($t\rightarrow 0$), 
to first order in $t$ 
\eqn\ten{
e^\phi~\sim~{\rm constant} + {\cal O}(t^2),
}
\eqn\eleven{
a~\sim~ 1 + {q\over {2\sqrt 3}}t + {\cal O}(t^2).}
In fig.2 and fig.3, we have ploted the scale 
factor and the dilaton as
functions of time for $\om =1$ and $q^2 =5$. 
So we notice that at late 
time, the effect of axion on the graviton-dilaton 
system is practically 
nil. However, at early time the axion seems to have significant 
influence on the behaviour of the universe.

To this end, we would also like to mention 
that, there is another set of 
solutions of \one\ - \three\ 
which would correspond to the $t <0$ branch. Those are given by:
\eqn\bfour{
\phi = -{1\over 2}{\rm ln}[{c(2\om +3)\over {4q^2}}{\rm sech}^2{
\sqrt c\tau\over 2}]}
\eqn\bfive{
a = [{4q^2\over {c(2\om +3)}}{\rm cosh}^2{\sqrt c\tau\over 2}]^{1\over 
4}  
e^{-{1\over 4}{\sqrt{c(2\om +3)\over 3}\tau}}}
and 
\eqn\bseven{
\bp = {1\over 4}[{\rm ln}\{{c(2\om +3)\over {4q^2}}{\rm sech}^2{
\sqrt c\tau\over 2}\} +\sqrt{3c(2\om +3)}\tau]}
with $t$ and $\tau$ are realated as in \six\ .

\bigskip

\noindent{\bf 3. Appendix.}

Here we briefly mention one way to solve \one\ - \three\ .

Define $\bal$ and $\bp$(the one introduced in the text to define 
dialton-time $\tau$) as 
\eqn\aone{\alpha = A \bal + B\bp,}
\eqn\atwo{\phi = C\bal + D\bp,}
where $A, B, C, D$ are 
\eqn\athree{A = {1\over {3\om +4}}\sqrt{{2\om +3}\over 3},
~~~B = -{{1+\om}\over{3\om +4}}}
\eqn\afour{C = {\sqrt 3\over{3\om +4}}\sqrt{{2\om +3}\over 3},
~~~D ={1\over {3\om +4}}.}
Further, defining 
\eqn\afive{Y = 3A\bal + D\bp,~~~X = D\bal + 3A\bp ,}
the equations can be reduced to 
\eqn\asix{Y^{\prime\prime} - {q^2\over {(2\om +3)}}e^{-2Y} =0,
~~~X^{\prime\prime} = 0, }
along with the constraint equation
\eqn\aseven{
(Y^\prime)^2 - (X^\prime)^2 + {q^2\over {2\om +3}}e^{-2Y} = 0.}
Here, in the above equations, prime 
denotes differentiation with respect
to $\tau$. Now, note that the first equation of \asix\ 
is just the Liouville like equation which 
can easily be integrated. By suitably choosing 
the constant of integrations
arising from \asix\ , the 
constraint equation \aseven\ can be 
made to satisfy.
\bigskip

\noindent {\bf Acknowledgements}: I thank E. Copeland 
for a very helpful
correspondence and pointing me out the reference \refs{\mw}.
I also thank J. Maharana and S. Panda 
for discussions on \refs{\veneziano}, \refs{\lidsey}. The work is 
supported by NSF Grant No. PHY-9411543. \vfill\eject
\listrefs
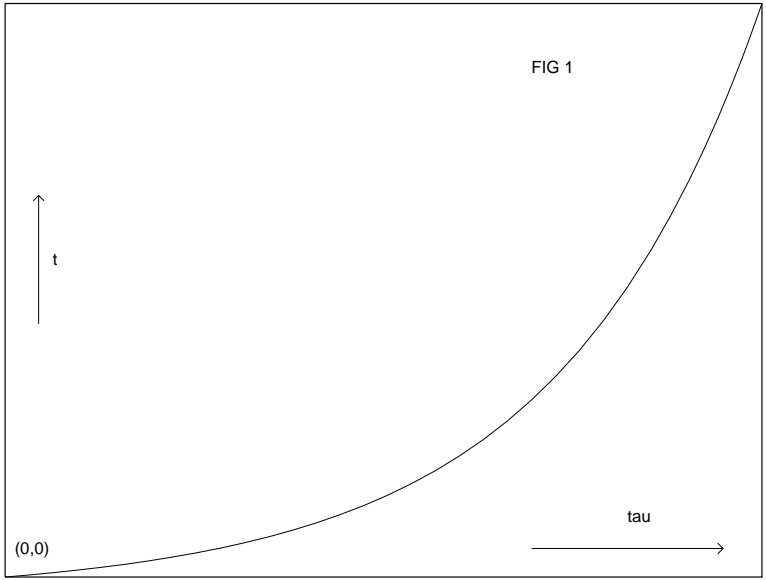
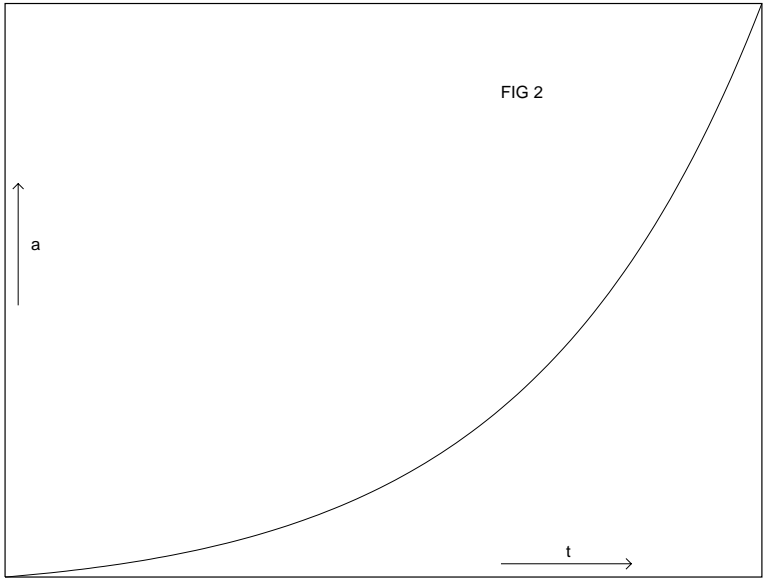
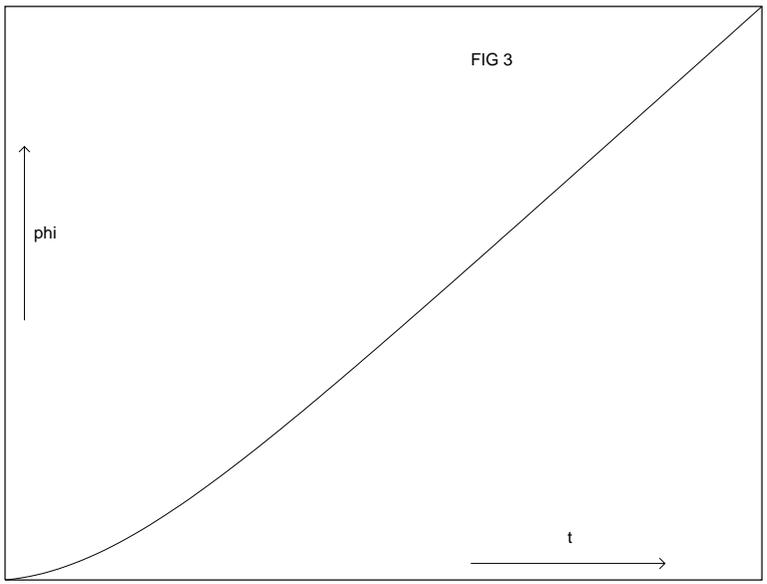
\bye